# Magneto-optical switching in microcavities based on a TGG defect sandwiched between periodic and disordered one-dimensional photonic structures

Ilka Kriegel [1], Francesco Scotognella [2,3,*]

[1]Department of Nanochemistry, Istituto Italiano di Tecnologia (IIT), via Morego, 30, 16163 Genova, Italy
[2]Dipartimento di Fisica, Politecnico di Milano, Piazza Leonardo da Vinci 32, 20133 Milano, Italy
[3]Center for Nano Science and Technology@PoliMi, Istituto Italiano di Tecnologia, Via Giovanni Pascoli, 70/3, 20133, Milan, Italy
*email address: francesco.scotognella@polimi.it

**Abstract**
The employment of magneto-optical materials to fabricate photonic crystals gives the unique opportunity to achieve optical tuning by applying a magnetic field. In this study we have simulated the transmission spectrum of a microcavity in which the Bragg reflectors are made with silica ($SiO_2$) and yttria ($Y_2O_3$) and the defect layer is made with TGG ($Tb_3Ga_5O_{12}$). We show that the application of an external magnetic field results in a tuning of the defect mode of the microcavity. In the simulations we have considered the wavelength dependence of the refractive indexes and the Verdet constants of the materials. A tuning of the defect mode of about 22 nm with a magnetic field of 5 T, at low temperature (8 K), is demonstrated. Furthermore, we discuss the possibility to tune a microcavity with disordered photonic structures as reflectors. In the presence of the magnetic field such microcavity shows a shift of resonances in a broad range of wavelengths. This study presents a method of contactless optical tuning.

**Keywords:** photonic crystals; magneto-optical properties; magnetic field.

**Introduction**
Photonic crystals are among the most interesting structures in optics, that give the possibility to transmit or reflect only certain energy ranges of the electromagnetic radiation. The energy region that is reflected by the photonic crystal is the so called photonic band gap [1–4]. A topic, which is attracting increasing attention, is the active tuning of the photonic band gap. The most simple external stimulus that can be employed for such tuning is the electric field, as described in a recent and exhaustive review article [5]. To report some examples of electro-optical tuning, electro-active polyferrocenylsilane based opals and inverse opals with a broad tunability have been reported [6,7]. A liquid crystal infiltrated one dimensional photonic crystal showed a band gap shift of 8 nm by applying a field of 8 V [8,9]. Moreover, with a silver nanoparticle/titania nanoparticle based one dimensional photonic crystal, a band gap shift of 10 nm with 10 V applied voltage has been achieved [10]. The employment of electric contacts is necessary to apply an external electric field,. To achieve a tuning of the photonic band gap without electric contacts, the application of a magnetic field is a promising way [11–14]. In fact, in recent years interesting works were reported on magnetophotonic crystals [15–18]. In this work, we focus on the effect of microcavities on the light transmission spectrum in presence of an external magnetic field using clockwise and counter-clockwise polarizations. We simulate the transmission spectra of a microcavity in which the Bragg reflectors are made alternating layers of silica ($SiO_2$) and yttria ($Y_2O_3$) and the defect layer is made of TGG ($Tb_3Ga_5O_{12}$). TGG is a widely investigated material showing interesting magnetic properties as a very high Verdet constant [19,20]. The wavelength dependent refractive indexes and Verdet constants of the materials has been judiciously considered in the simulation. We show the possibility to achieve the tuning of the defect mode in the photonic band gap with the employment of the magnetic field. We can show a tuning of the defect mode of 22 nm with of 5 T, at low temperature. We also show the possibility to tune a microcavity with disordered Bragg reflectors, with a shift of resonances in a broad range of wavelengths.

**Outline of the method**
For a careful simulation of the light transmission through the microcavities, we need to take into account the dispersion of the refractive index and the Verdet constant of the different materials employed. With the magnetic field $\vec{B}$ parallel to the direction of propagation of light, the difference between the refractive indexes for left and right circularly polarized (clockwise and counter-clockwise, respectively) beams that propagate through the medium can be expressed in the following way [11]:

$$n_{R,L}(\lambda) = \sqrt{\varepsilon\mu} \pm \frac{V\vec{B}\lambda}{2\pi} \quad (1)$$

where $\varepsilon$ is the dielectric permittivity and $\mu$ is the magnetic permittivity. $V$ is the Verdet constant and $\lambda$ is the wavelength. If we assume, for the materials employed in this work, that $\sqrt{\varepsilon\mu} = \sqrt{\varepsilon} = n(\lambda)$ [i.e., $\mu \approx 1$, also for TGG [21]], the first term is the wavelength dependent index of refraction. Moreover, we need to consider also the wavelength dependency the Verdet constant $V$, such that the expression becomes:

$$n_{R,L}(\lambda) = n(\lambda) \pm \frac{V(\lambda)\vec{B}\lambda}{2\pi} \quad (2)$$

In the structure studied we used three materials: TGG ($Tb_3Ga_5O_{12}$), silica ($SiO_2$) and yttria ($Y_2O_3$).

*Refractive index and Verdet constant of TGG*
For the dispersion of the refractive index of TGG we fit the experimental data from [22] with a single absorption resonance Sellmeier equation:

$$n^2_{TGG}(\lambda) - 1 = \frac{2.742\lambda^2}{\lambda^2 - 0.01743} \quad (3)$$

where the wavelength is in micrometers. The $R^2$ of the fit is > 0.999.
A very detailed study on TGG reports the wavelength and temperature dependent Verdet constant [20]

$$V(\lambda, T) = +\frac{175\lambda_0^2}{\lambda^2 - \lambda_0^2} - \frac{(32\times10^4)\lambda_0^2}{(T-T_W)(\lambda^2 - \lambda_0^2)} - \frac{1714}{T-T_W} \quad (4)$$

where the wavelength is in nm. Such Verdet constant is referred to the ceramic material K4 (by Konoshima Chemical, Co. Ltd.). $\lambda_0$, 237 nm for K4, is the wavelength of the absorption resonance, and $T_W$, 7.6 K for K4, is the Curie-Weiss temperature [20].

*Refractive index and Verdet constant of Silica*
For silica we used the Sellmeier equation as reported in Ref. [23]:

$$n^2_{SiO_2}(\lambda) - 1 = \frac{0.6961663\lambda^2}{\lambda^2 - 0.0684043^2} + \frac{0.4079426\lambda^2}{\lambda^2 - 0.1162414^2} + \frac{0.8974794\lambda^2}{\lambda^2 - 9.896161^2} \quad (5)$$

For the Verdet constant we fit the data from Refs. [24–26] to achieve the following expression:

$$V(\lambda) = \frac{1.45}{\lambda^2 - 0.1252^2} \quad (6)$$

*Refractive index ad Verdet constant of Yttria*
For Yttria we used the Sellmeier equation as reported in Ref. [27]:

$$n^2_{Y_2O_3}(\lambda) - 1 = \frac{0.2578\lambda^2}{\lambda^2 - 0.1387^2} + \frac{3.935\lambda^2}{\lambda^2 - 22.936^2} \quad (7)$$

where the wavelength is in micrometers. The $R^2$ of the fit is > 0.999.

For the Verdet constant of Yttria we fit the data from Ref. [28] by using a power law equation $f(\lambda) = \frac{a}{\lambda^b} + c$:

$$V(\lambda) = \frac{0.007011}{\lambda^{3.08}} + 1.297 \tag{8}$$

where the wavelength is in micrometers. The $R^2$ of the fit is > 0.999.

We used the refractive indexes as expressed in Equation 2. The first term of the equation is the Sellmeier equation, while the second term includes the wavelength dependent Verdet constant (wavelength and temperature dependent for TGG). Such refractive indexes have been employed in the transfer matrix method to simulate the transmission spectra [29–31]. We have performed the simulation with a spectral resolution of 1 picometer.

**Results and Discussion**

In Figure 1 we show the scheme of the studied microcavity in which a TGG defect is embedded between two Bragg reflectors. Each Bragg reflector consists of 10 silica/yttria bilayers. In this simulation the thickness of the silica layers is 0.12 μm, the one of the yttria layers is 0.07 μm, while the thickness of TGG is 0.69 μm, a value similar to the one used in another interesting simulation reported in Ref. [32].

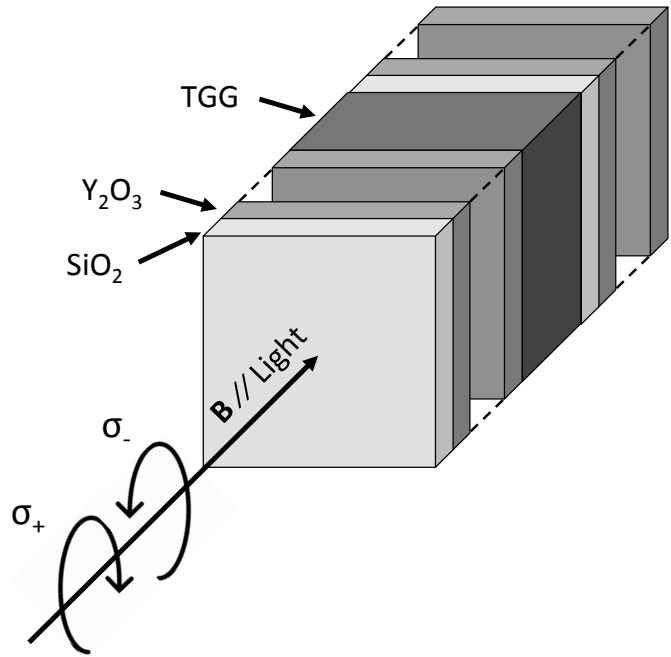

**Figure 1.** Scheme of the $(SiO_2/Y_2O_3)_{10}/TGG/(SiO_2/Y_2O_3)_{10}$ microcavity, i.e. a TGG layer sandwiched between two Bragg reflectors made by 10 bilayers of $SiO_2/Y_2O_3$.

In Figure 2 we show the transmission spectrum of the microcavity for clockwise and counter-clockwise polarized light. The magnetic field applied is 5 Tesla and the temperature of the experiment is 8 K. We simulate the transmission spectrum at wavelengths longer than 0.52 μm to avoid the absorption resonance of TGG at 0.488 μm [19].

We observed that the defect mode of the microcavity is sensitive to the polarization of light in presence of the magnetic field. The defect mode central wavelength for the clockwise polarization is red-shifted by about 22 nm with respect to the defect mode for the counter-clockwise polarization. This also means a shift of about 11 nm of the defect mode by turning on/off the magnetic field (Figure 2b).

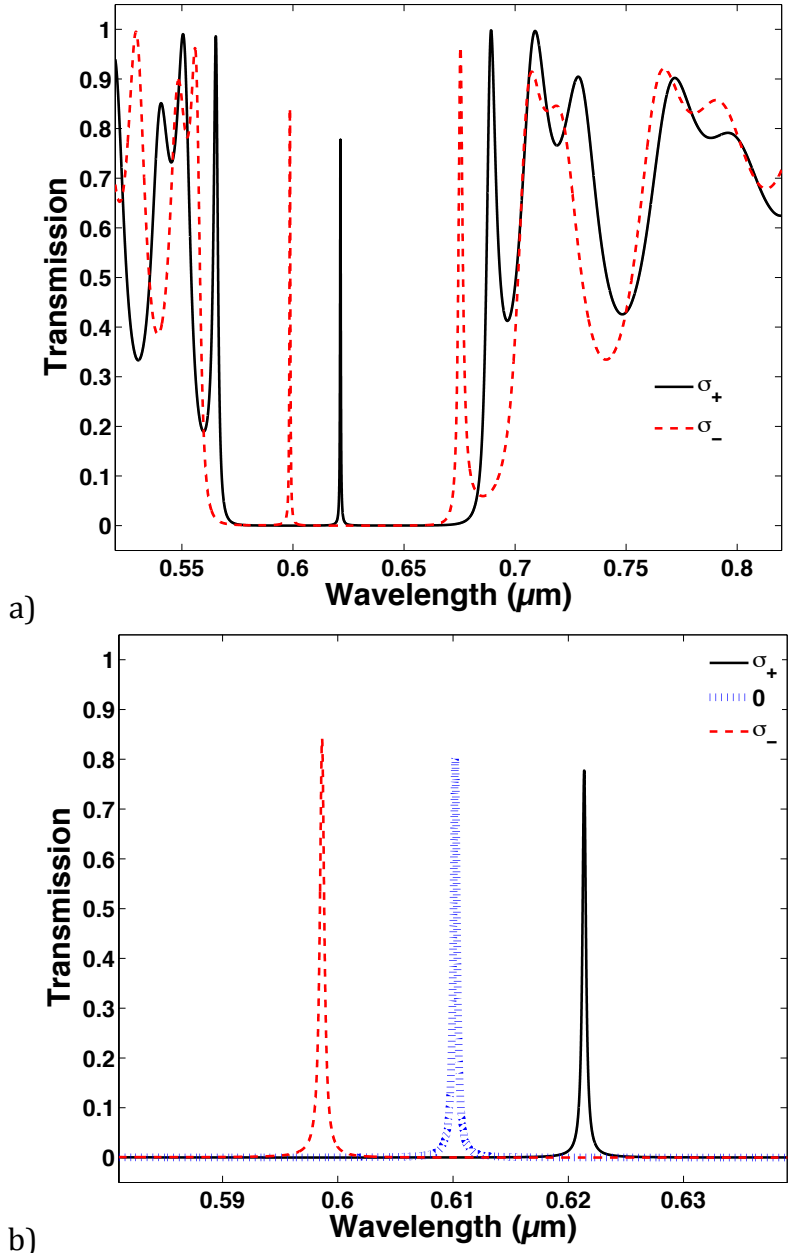

**Figure 2.** (a) Transmission spectrum of the $(SiO_2/Y_2O_3)_{10}/TGG/(SiO_2/Y_2O_3)_{10}$ microcavity for the clockwise $\sigma_+$ (black solid line) and the counter-clockwise $\sigma_-$ (red dashed line) polarizations. (b) Zoom in the wavelength range of the bad gap defect mode of the microcavity for the clockwise $\sigma_+$ (black solid line) and the counter-clockwise $\sigma_-$ (red dashed line) polarizations and for the microcavity without the applied magnetic field ($\vec{B} = 0$, blue short-dashed line).

Figure 3 depicts the spectral position of the band gap defect as a function of the applied magnetic field and temperature for the two polarizations. Such intensities of the magnetic field, and values of temperature, can be applied with a setup as, for example, the one described in Ref. [33]. In Figure 3b, where a step of 1 K has been used, we clearly see the non-monotonic trend in agreement with equation 4 (the stars in Figure 3b depict the spectral positions very close to $T_W$ if a temperature step of 0.1 K is feasible).

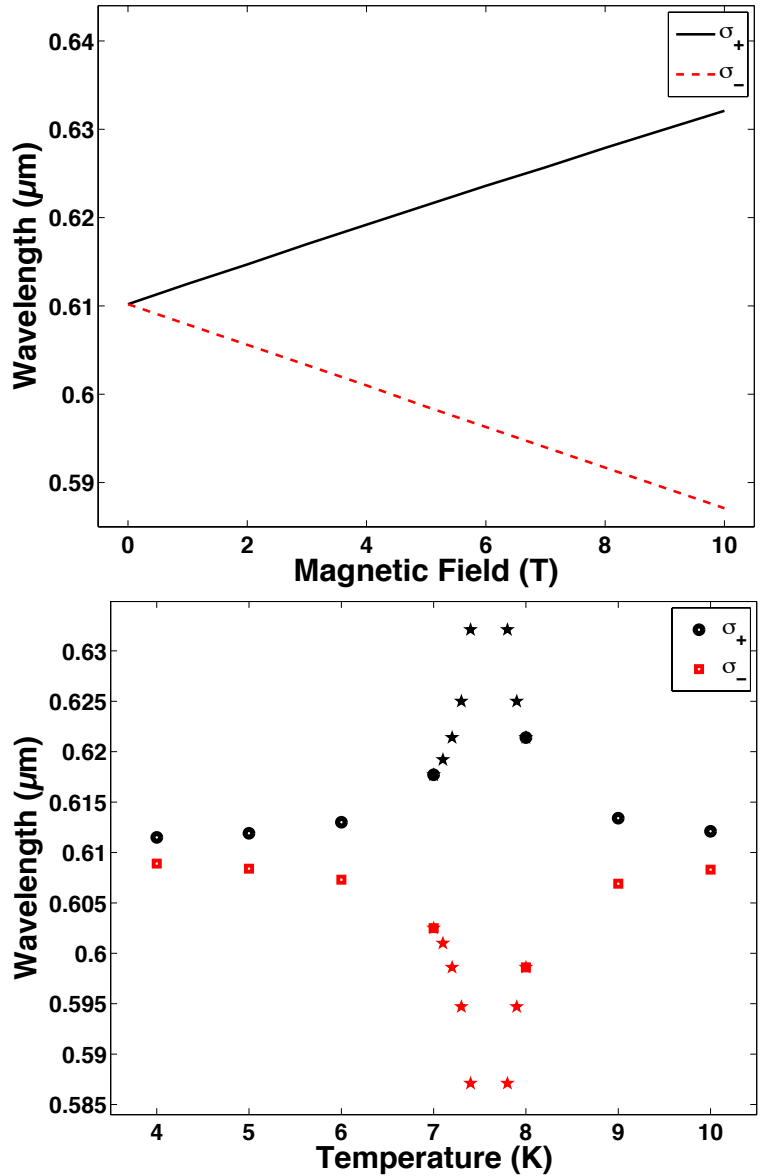

**Figure 3.** Spectral position of the band gap defect, for the clockwise $\sigma_+$ (black circles) and the counter-clockwise $\sigma_-$ (red squares) polarizations, (a) as a function of the applied magnetic field (temperature $T$ = 8K), and (b) as a function of the temperature (with an applied magnetic field of 5 T). The stars in (b) depict the spectral positions at temperatures very close to $T_W$.

In Figure 4 we show the shift of several resonances (e.g. from 0.560 to 0.545 μm, and from 0.629 to 0.612 μm) in a microcavity in which TGG is embedded between two disordered one-dimensional photonic structures made of $SiO_2$ and $Y_2O_3$. The disordered structures are made with 20 layers each, where each layer has 50% probability to either $SiO_2$ or $Y_2O_3$. For details on disordered one dimensional photonic structures the reader is referred to reference [34,35]. Notably, in this configuration a wavelength shift on several resonances is observed, with respect to the effect on the narrow line width cavity mode presented in Figure 2. We believe the difference here is a result of the disorder introduced in the photonic structures that embed the TGG defect. The latter example is an interesting way to tune the broad transmission properties of a photonic structure in a contactless way by applying a magnetic field, useful for example for a tunable random laser.

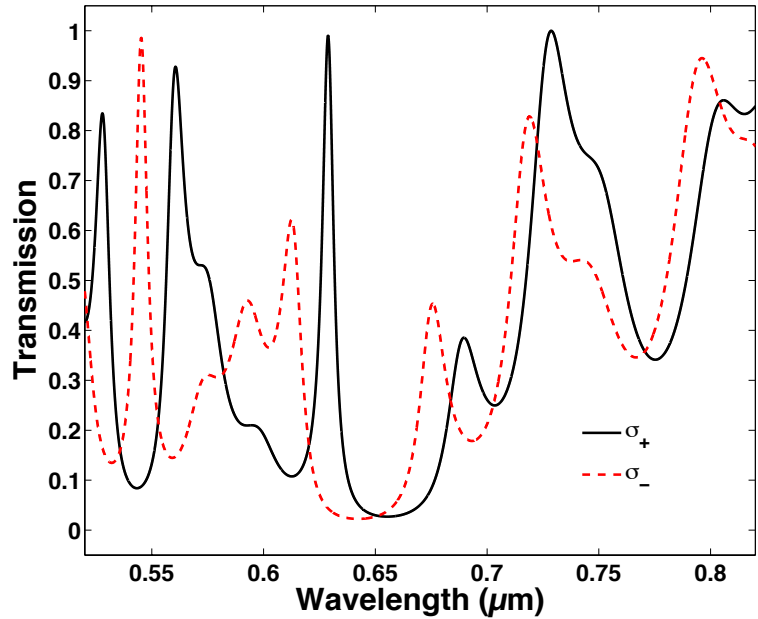

**Figure 4.** Transmission spectrum of a microcavity in which TGG is embedded between two disordered one-dimensional photonic structures made of $SiO_2$ and $Y_2O_3$. We show the clockwise $\sigma_+$ (black solid line) and the counter-clockwise $\sigma_-$ (red dashed line) polarizations.

**Conclusion**

In this study we simulated the transmission spectrum of two types of microcavities employing as materials silica ($SiO_2$) and yttria ($Y_2O_3$) and TGG ($Tb_3Ga_5O_{12}$) as defect layer. The first microcavity sandwiches the TGG defect between two periodic Bragg reflectors, the second one sandwiches the TGG defect between two one-dimensional disordered structures. We show that the application of an external magnetic field results in a tuning of the defect mode of the microcavities. Although this effect is a well known physical phenomenon [11], only the careful studies of the wavelength and temperature dependent Verdet constant of TGG [19,20] allow the desing of photonics structures embedding TGG. In fact, in the simulations we have considered the dispersions of the refractive indexes and the Verdet constants of the materials. A tuning of the defect mode of about 22 nm with of 5 T, at low temperature (8 K), is demonstrated. The magnetic field and temperature conditions can be achieved by placing the sample in the core of a superconducting solenoid and in contact with helium gas (example of a proper setup is reported in Ref. [33]). With respect to the periodic microcavity, the one employing two disordered structures shows a variety of tunable transmission peaks in the studied spectral range. A fabrication of the described structures can be pursued with techniques as sputtering [36] and pulsed laser deposition [37].

# References

[1]	E. Yablonovitch, Inhibited Spontaneous Emission in Solid-State Physics and Electronics, Phys. Rev. Lett. 58 (1987) 2059–2062. doi:10.1103/PhysRevLett.58.2059.
[2]	S. John, Strong localization of photons in certain disordered dielectric superlattices, Phys. Rev. Lett. 58 (1987) 2486–2489. doi:10.1103/PhysRevLett.58.2486.
[3]	J.D. Joannopoulos, Photonic crystals: molding the flow of light, Princeton University Press, Princeton, 2008.
[4]	K. Sakoda, Optical Properties of Photonic Crystals, Springer Science & Business Media, 2004.
[5]	L. Nucara, F. Greco, V. Mattoli, Electrically responsive photonic crystals: a review, J Mater Chem C. 3 (2015) 8449–8467. doi:10.1039/C5TC00773A.
[6]	A.C. Arsenault, D.P. Puzzo, I. Manners, G.A. Ozin, Photonic-crystal full-colour displays, Nat. Photonics. 1 (2007) 468–472. doi:10.1038/nphoton.2007.140.
[7]	D.P. Puzzo, A.C. Arsenault, I. Manners, G.A. Ozin, Electroactive Inverse Opal: A Single Material for All Colors, Angew. Chem. Int. Ed. 48 (2009) 943–947. doi:10.1002/anie.200804391.
[8]	L. Criante, F. Scotognella, Low-Voltage Tuning in a Nanoparticle/Liquid Crystal Photonic Structure, J. Phys. Chem. C. 116 (2012) 21572–21576. doi:10.1021/jp309061r.
[9]	L. Criante, F. Scotognella, Infiltration of E7 Liquid Crystal in a Nanoparticle-Based Multilayer Photonic Crystal: Fabrication and Electro-optical Characterization, Mol. Cryst. Liq. Cryst. 572 (2013) 31–39. doi:10.1080/15421406.2012.763207.
[10]	E. Aluicio-Sarduy, S. Callegari, D.G. Figueroa del Valle, A. Desii, I. Kriegel, F. Scotognella, Electric field induced structural colour tuning of a silver/titanium dioxide nanoparticle one-dimensional photonic crystal, Beilstein J. Nanotechnol. 7 (2016) 1404–1410. doi:10.3762/bjnano.7.131.
[11]	A. Baev, P.N. Prasad, H. Ågren, M. Samoć, M. Wegener, Metaphotonics: An emerging field with opportunities and challenges, Phys. Rep. 594 (2015) 1–60. doi:10.1016/j.physrep.2015.07.002.
[12]	D. Bossini, V.I. Belotelov, A.K. Zvezdin, A.N. Kalish, A.V. Kimel, Magnetoplasmonics and Femtosecond Optomagnetism at the Nanoscale, ACS Photonics. 3 (2016) 1385–1400. doi:10.1021/acsphotonics.6b00107.
[13]	S. Pu, S. Dong, J. Huang, Tunable slow light based on magnetic-fluid-infiltrated photonic crystal waveguides, J. Opt. 16 (2014) 045102. doi:10.1088/2040-8978/16/4/045102.
[14]	D. Su, S. Pu, L. Mao, Z. Wang, K. Qian, A Photonic Crystal Magnetic Field Sensor Using a Shoulder-Coupled Resonant Cavity Infiltrated with Magnetic Fluid, Sensors. 16 (2016) 2157. doi:10.3390/s16122157.
[15]	M. Inoue, K. Arai, T. Fujii, M. Abe, Magneto-optical properties of one-dimensional photonic crystals composed of magnetic and dielectric layers, J. Appl. Phys. 83 (1998) 6768. doi:10.1063/1.367789.
[16]	V.I. Belotelov, A.K. Zvezdin, Magneto-optical properties of photonic crystals, J. Opt. Soc. Am. B. 22 (2005) 286. doi:10.1364/JOSAB.22.000286.
[17]	M. Inoue, R. Fujikawa, A. Baryshev, A. Khanikaev, P.B. Lim, H. Uchida, O. Aktsipetrov, A. Fedyanin, T. Murzina, A. Granovsky, Magnetophotonic crystals, J. Phys. Appl. Phys. 39 (2006) R151. doi:10.1088/0022-3727/39/8/R01.
[18]	A.M. Grishin, S.I. Khartsev, All-Garnet Magneto-Optical Photonic Crystals, J. Magn. Soc. Jpn. 32 (2008) 140–145. doi:10.3379/msjmag.32.140.
[19]	O. Slezak, R. Yasuhara, A. Lucianetti, T. Mocek, Wavelength dependence of magneto-optic properties of terbium gallium garnet ceramics, Opt. Express. 23 (2015) 13641. doi:10.1364/OE.23.013641.
[20]	O. Slezák, R. Yasuhara, A. Lucianetti, T. Mocek, Temperature-wavelength dependence of

terbium gallium garnet ceramics Verdet constant, Opt. Mater. Express. 6 (2016) 3683. doi:10.1364/OME.6.003683.

[21]    U. Löw, S. Zvyagin, M. Ozerov, U. Schaufuss, V. Kataev, B. Wolf, B. Lüthi, Magnetization, magnetic susceptibility and ESR in Tb3Ga5O12, Eur. Phys. J. B. 86 (2013). doi:10.1140/epjb/e2012-30993-0.

[22]    R. Yasuhara, H. Nozawa, T. Yanagitani, S. Motokoshi, J. Kawanaka, Temperature dependence of thermo-optic effects of single-crystal and ceramic TGG, Opt. Express. 21 (2013) 31443. doi:10.1364/OE.21.031443.

[23]    I.H. Malitson, Interspecimen Comparison of the Refractive Index of Fused Silica, J. Opt. Soc. Am. 55 (1965) 1205–1208. doi:10.1364/JOSA.55.001205.

[24]    S. Fujioka, Z. Zhang, K. Ishihara, K. Shigemori, Y. Hironaka, T. Johzaki, A. Sunahara, N. Yamamoto, H. Nakashima, T. Watanabe, H. Shiraga, H. Nishimura, H. Azechi, Kilotesla Magnetic Field due to a Capacitor-Coil Target Driven by High Power Laser, Sci. Rep. 3 (2013). doi:10.1038/srep01170.

[25]    F. Mitschke, Fiber optics: physics and technology, Springer, Heidelberg ; New York, 2009.

[26]    G.W.C. Kaye, T.H. Laby, Tables of Physical and Chemical Constants and Some Mathematical Functions (16th edition), 1995.

[27]    Y. Nigara, Measurement of the Optical Constants of Yttrium Oxide, Jpn. J. Appl. Phys. 7 (1968) 404–408. doi:10.1143/JJAP.7.404.

[28]    Kruk, Andrzej, Wajler, Anna, Mrozek, Mariusz, Zych, Lukasz, Gawlik, Wojciech, Brylewski, Tomasz, Transparent yttrium oxide ceramics as potential optical isolator materials, Opt. Appl. 45 (2015) 585–594. doi:10.5277/oa150413.

[29]    M. Born, E. Wolf, Principles of Optics: Electromagnetic Theory of Propagation, Interference and Diffraction of Light, Cambridge University Press, 2000.

[30]    M. Bellingeri, I. Kriegel, F. Scotognella, One dimensional disordered photonic structures characterized by uniform distributions of clusters, Opt. Mater. 39 (2015) 235–238. doi:10.1016/j.optmat.2014.11.033.

[31]    X. Xiao, W. Wenjun, L. Shuhong, Z. Wanquan, Z. Dong, D. Qianqian, G. Xuexi, Z. Bingyuan, Investigation of defect modes with Al2O3 and TiO2 in one-dimensional photonic crystals, Opt. - Int. J. Light Electron Opt. 127 (2016) 135–138. doi:10.1016/j.ijleo.2015.10.005.

[32]    T. Goto, R. Isogai, M. Inoue, Para-magneto- and electro-optic microcavities for blue wavelength modulation, Opt. Express. 21 (2013) 19648. doi:10.1364/OE.21.019648.

[33]    F. Liu, L. Biadala, A.V. Rodina, D.R. Yakovlev, D. Dunker, C. Javaux, J.-P. Hermier, A.L. Efros, B. Dubertret, M. Bayer, Spin dynamics of negatively charged excitons in CdSe/CdS colloidal nanocrystals, Phys. Rev. B. 88 (2013). doi:10.1103/PhysRevB.88.035302.

[34]    D.S. Wiersma, R. Sapienza, S. Mujumdar, M. Colocci, M. Ghulinyan, L. Pavesi, Optics of nanostructured dielectrics, J. Opt. Pure Appl. Opt. 7 (2005) S190. doi:10.1088/1464-4258/7/2/025.

[35]    D.S. Wiersma, Disordered photonics, Nat. Photonics. 7 (2013) 188–196. doi:10.1038/nphoton.2013.29.

[36]    S. Valligatla, A. Chiasera, S. Varas, N. Bazzanella, D.N. Rao, G.C. Righini, M. Ferrari, High quality factor 1-D Er3+-activated dielectric microcavity fabricated by RF-sputtering, Opt. Express. 20 (2012) 21214–21222. doi:10.1364/OE.20.021214.

[37]    L. Passoni, L. Criante, F. Fumagalli, F. Scotognella, G. Lanzani, F. Di Fonzo, Self-Assembled Hierarchical Nanostructures for High-Efficiency Porous Photonic Crystals, ACS Nano. 8 (2014) 12167–12174. doi:10.1021/nn5037202.